\documentclass[prd,aps,showpacs,notitlepage,nofootinbib,tightenlines]{revtex4}  
\usepackage{mathrsfs}
\usepackage{amsmath}
\usepackage{amssymb}
\usepackage{epsfig}
\usepackage{graphicx}
\usepackage{booktabs}
\usepackage{multirow}
\usepackage{subfigure}

\newcommand{\beq}{\begin{eqnarray}}
\newcommand{\eeq}{\end{eqnarray}}
\newcommand{\non}{\nonumber\\ }
\begin{document}
\title{Branching ratios, $CP$ asymmetries and polarizations of $B\rightarrow \psi(2S) V$ decays }
\author{Zhou Rui$^1$}\email{jindui1127@126.com}
\author{Ya Li$^2$}\email{liyakelly@163.com}
\author{ Zhen-Jun Xiao$^2$}\email{xiaozhenjun@njnu.edu.cn}
\affiliation{$^1$ College of Sciences, North China University of Science and Technology,
                          Tangshan 063009,  China}
\affiliation{$^2$ Department of Physics and Institute of Theoretical Physics,
                          Nanjing Normal University, Nanjing 210023, Jiangsu, China}
\date{\today}

\begin{abstract}
We analyzed the nonleptonic decays $B/B_s\to \psi(2S) V $ with $V=(\rho, \omega, K^{*}, \phi)$
by employing the perturbative QCD (PQCD) factorization approach.
Here the branching ratios, the $CP$ asymmetries and the complete set of polarization
observables are investigated systematically.
Besides the traditional contributions from  the factorizable and nonfactorizable diagrams
at the leading order, the next-to-leading order (NLO) vertex corrections could also provide considerable contributions.
The  PQCD predictions for the branching ratios of the  $B_{(s)}\to \psi(2S)K^{*},  \psi(2S)\phi$
decays are consistent with the measured values within errors.
As for $B\to \psi(2S) \rho, \psi(2S) \omega$ decays, the branching ratios can
reach the order of $10^{-5}$
and could be measured in the LHCb and Belle-II experiments.
The numerical results show that the direct $CP$ asymmetries of the considered decays  are very small.
Thus the observation of any large direct $CP$ asymmetry for these decays will be a signal for new physics.
The mixing induced $CP$ asymmetries in the neutral modes are very close to $\sin 2\beta_{(s)}$,
which suggests that these channels can give a cross-check on the measurement of the    Cabbibo-Kobayashi-Maskawa (CKM) angle $\beta$ and $\beta_s$.
We found that the longitudinal polarization fractions $f_0$ are suppressed to $\sim 50\%$
due to the large nonfactorizable contributions.
The magnitudes and phases of the two transverse amplitudes $\mathcal {A}_{\parallel}$ and $\mathcal {A}_{\perp}$ are roughly
equal, which is an indication for the approximate light quark helicity conservation in these decays.
The overall polarization observables of $B\to \psi(2S) K^{*0}$ and $B_s\to \psi(2S) \phi$ channels are also in
good agreement with the experimental measurements as reported by LHCb and BaBar.
Other results can also be tested by the LHCb and Belle-II experiments.
\end{abstract}



\maketitle

\section{Introduction}

Studies of the decays of $B$ mesons into  a charmonium meson plus a light vector meson
 contribute a lot to our knowledge about the $CP$ violation  and mixing in the $B$ meson system \cite{npb19385},
 and also provide a particularly important place to look for the physics
beyond the standard model (SM). For example, the mode $B^0_s\rightarrow J/\psi \phi $ is the  so-called ``golden mode" for measuring $\beta_s$,
 which is extracted from the angular analysis of the time-dependent differential decay rate
 \cite{prd85072002,prd85032006,prl108101803}.
 The counterpart phase $\beta$ in the $B$ meson system can also be extracted
 in  $B^0\rightarrow J/\psi K^* $ decay \cite{epjc743026}.
 The decay $B^0\rightarrow J/\psi \phi $,
which would proceed mainly via a Cabibbo-suppressed and color-suppressed transition ($\bar{b}d\rightarrow \bar{c}c\bar{d}d$)  with rescattering of $d\bar{d}$
into $s\bar{s}$, provide useful information for understanding rescattering mechanisms \cite{prd78011106,prd88072005,prd91012003}.
In addition, combining the decays $B^0\rightarrow J/\psi \phi $ and  $B^0\rightarrow J/\psi \omega $   can be helpful to study the $\omega-\phi$ mixing \cite{plb666185}.
These decays are dominated by tree diagrams and the contributions from penguin diagrams are expected to be small.
With continuously increasing high-precision measurements, the penguin effects, which play an important role in
the extraction of  the above  phases, can be measured by means of an analysis of the angular distribution of
$B^0\rightarrow J/\psi \rho^0 $ \cite{prd60073008} and $B^0_s\rightarrow J/\psi \bar{K}^{*0} $ \cite{prd79014005}.

In the framework of SM, these decay modes are induced by  transitions   $b\to q c \bar{c}$ with $q=d,s$.
In principle, any mode involving various excitations of the $c\bar{c}$ mesons such as $B\to \psi(2S) V$ decays 
could be an alternative to  that for $J/\psi$ analogues, and give additional and complementary information.
Experimentally,
the $\psi(2S)$ meson can be  reconstructed in the decay channels $\psi(2S)\rightarrow \mu^+\mu^-$ and
$\psi(2S)\rightarrow J/\psi \pi^+\pi^-$, with the $J/\psi$ meson decaying into two muons \cite{epjc722100}.
Nowadays, several experimental Collaborations 
have measured the decays $B_s^0\rightarrow \psi(2S) \phi$ \cite{plb762253},
$B\rightarrow \psi(2S) K^*(892)$ \cite{prd58072001,prd63031103,prd65032001,prl94141801,prd76031102},
$B^0 \rightarrow \psi(2S) \pi^0$ \cite{prd93031101}, $B^+ \rightarrow \psi(2S) \pi^+$ \cite{prd78051104},
$B_s \rightarrow \psi(2S) \eta^{(')}$ \cite{npb871403,jhep01024}.
Some relative ratios of the branching ratios for $B$ meson decays into $\psi(2S)$ and $J/\psi$ mesons
are also measured by  several experiments \cite{prd85091105,epjc722118,prd79111102,prl96231801}.

On the theory side, these $B \to \psi(2S)V$ modes do have some special properties.
Since there are three possible values of the total angular momentum with different $CP$
eigenvalues ($L=1$ is odd, while $L=0,2$ are even),
the angular analysis is needed to   separate the contributions from the $CP$-even
and $CP$-odd partial waves. Therefore the final state can be decomposed into three helicity amplitudes
(one longitudinal and two transverse components ).
The information  about the phases of the transverse decay amplitudes provides a test of the factorization
hypothesis \cite{zpc29637}.
Furthermore, $B \rightarrow \psi(2S)V$ are the color-suppressed modes and therefore a significant
impact of nonfactorizable contributions  is expected. Both improvements in the accuracy of
the experimental measurements and the observation of new modes, especially involving    $\psi(2S)$
in the final states, can be helpful in understanding the role of any nonfactorizable
corrections \cite{prd59092004,prd68034004,prd71114008} and differentiating various theory approaches.
Nowadays there exist several theoretical approaches as described in
Refs.~\cite{prd65094023,prd71016007,prd73014016,prd79014030,prd83094027,prd90094006,jhep03145,prl115061802,prd86053008,jhep03009,
 epjc722229,prd89097503,prd86011501,prl111062001,epjc711798,prd87074021}
which shed more light on the S-wave ground state charmonium decays of $B$ mesons.
The nonleptonic $B$ decays with radially excited charmonium mesons in final state, however, have received less
attention in the literature.

Based on  the $k_T$ factorization theorem,  The perturbative QCD (PQCD) approach \cite{pqcd1,pqcd2} is suitable for
describing different types of heavy hadron decays.
After including the parton transverse momentum $k_T$, which is not negligible in the end-point region, both factorizable
and  nonfactorizable  contributions are  calculable without  endpoint singularity.
The Sudakov resummation has also been introduced to suppress the long-distance contributions effectively.
Therefore, the PQCD approach is a self-consistent framework and has a good predictive power.
In our previous works \cite{zhourui1,zhourui2,zhourui3},  the semi-leptonic, two-body and  three-body non-leptonic
decays of the $B_c(B)$ mesons to $\psi(2S)$ are studied in the PQCD framework.
Here, furthermore, we will extend our previous analysis to the $B \rightarrow \psi(2S)V$ decays.
In a recent work \cite{prd89094010},  The authors  applied the PQCD approach to study $B\rightarrow J/\psi V$
decays and also obtained the theoretical predictions in good agreement with currently available data.
Therefore we have good reasons to believe that it is appropriate to analyze $B\rightarrow \psi(2S) V$ in this framework.

This work is organized as follows. In Sec. \ref{sec:f-work} we present some basic formulas such as the effective
Hamiltonian and kinetic conventions, then briefly review the pQCD approach.
The technical formulas of the calculation and the nonperturbative meson wave functions
are summarised in Appendix \ref{sec:ampulitude} and \ref{sec:wave}, respectively.
Section \ref{sec:results} devoted to numerical calculation and discussion.
Our conclusions are left for Sec. \ref{sec:conlusion}.

\section{ANALYTIC FORMULAS and  perturbative calculations}\label{sec:f-work}

For nonleptonic charmonium $B$ decays, both the tree operators and the penguin operators
of the standard effective weak Hamiltonian contribute, which is given by \cite{rmp681125}
\begin{eqnarray}\label{eq:operator}
\mathcal{H}_{eff}=\frac{G_F}{\sqrt{2}}\{\xi_c[C_1(\mu)O^c_1(\mu)+C_2(\mu)O^c_2(\mu)]
-\xi_t\sum_{i=3}^{10}C_i(\mu)O_i(\mu)\},
\end{eqnarray}
with the CKM matrix element $\xi_{c(t)}=V^*_{c(t)b}V_{c(t)q}$.
$O_i(\mu)$ and $C_i(\mu)$ are the effective four
quark operators and their QCD corrected Wilson coefficients at the renormalization scale $\mu$, respectively.
Their explicit form  can be found in Ref. \cite{rmp681125}.

\begin{figure}[tbp]
\centerline{\epsfxsize=9cm \epsffile{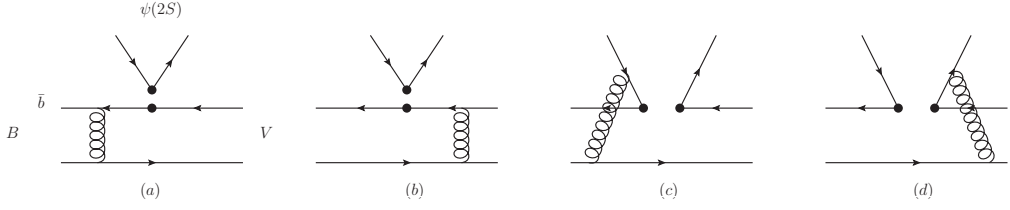}}
\vspace{-8cm}
\caption{The typical leading-order Feynman diagrams for the decay  $B \to \psi(2S)V$.
(a,b) The factorizable  diagrams,  and (c,d) the nonfactorizable diagrams.}
\label{fig:femy}
\end{figure}

At quark level, when the $\bar{b}\rightarrow \bar{q}c\bar{c}$ decay occurs through the
four quark operators, a $c\bar{c}$ state $\psi(2S)$ is created
while the other light anti-quark $\bar{q}$ is  flying away.
Since the heavy $b$ quark in $B$ meson carry most of the energy of $B$ meson, the spectator quark of the $B$ meson  is soft.
A hard gluon is exchanged so that the spectator quark gets energy from the
four quark operator and then form a fast moving vector meson with its partner anti-quark.
This makes the perturbative calculations into a six-quark interaction, which involves the four quark operator and
the spectator quark connected by a hard gluon. The relevant Feynman diagrams are shown in Fig.~\ref{fig:femy}.

In the PQCD approach, the decay amplitudes  are expressed as the convolution of the hard kernels $H$ with the
relevant meson wave functions  $\Phi_i$
\begin{eqnarray}
\label{eq:ampu}
\mathcal{A}(B\rightarrow \psi(2S) V) = \int d^4k_1d^4k_2d^4k_3 Tr[C(t)\Phi_B(k_1)\Phi_{\phi(2S)}(k_2)\Psi_V(k_3)
H(k_1,k_2,k_3,t)],
\end{eqnarray}
where $k_i$ are the momentum of the quark in each meson, and ``Tr" denotes the trace over all Dirac structure and
color indices.
$C(t)$ is the short distance Wilson coefficients  at the  hard-scale $t$.
The meson wave functions $\Phi$, including all nonperturbative components  in the $k_T$ factorization, can be
extracted from experimental data or other nonperturbative methods.
The hard kernel $H(k_i,t)$ describes the four quark operator and the spectator quark connected
by a hard gluon, which can be perturbatively calculated including all possible Feynman diagrams without endpoint singularity.
In the following, we start to compute the decay amplitudes of $B\rightarrow \psi(2S) V$ decay.

We will work in the $B$ meson rest frame and employ the light-cone coordinates for momentum variables.
The $B$ meson momentum $P_1$, the $\psi(2S)$ meson momentum $P_2$, the vector meson momentum $P_3$ and
the quark momenta $k_i$ in each meson are chosen  as
\begin{eqnarray}
 P_1&=&\frac{M}{\sqrt{2}}(1,1,\textbf{0}_{\rm T}),\quad P_2=\frac{M}{\sqrt{2}}(1-r_v^2,r^2,\textbf{0}_{\rm T}),\quad  P_3=\frac{M}{\sqrt{2}}(r_v^2,1-r^2,\textbf{0}_{\rm T}),\nonumber\\
  k_1&=&(\frac{M}{\sqrt{2}}x_1,0,\textbf{k}_{\rm 1T}),\quad k_2=(\frac{M}{\sqrt{2}}x_2(1-r_v^2),\frac{M}{\sqrt{2}}x_2r^2,\textbf{k}_{2\rm T}),\quad  k_3=(\frac{M}{\sqrt{2}}x_3r_v^2,\frac{M}{\sqrt{2}}x_3(1-r^2),\textbf{k}_{\rm 3T}),
\end{eqnarray}
with the mass ratio $r_{(v)}=m_{\psi(2S)}(m_V)/M$ and $m_{\psi(2S)},m_V,M$ are the masses of the charmonium, vector meson and $B$ meson, respectively.
The $k_{iT}$, $x_i$ represent the transverse momentum and longitudinal
momentum fraction of the quark inside the meson. Since the final state consists of two spin-1 particles,
to extract the helicity amplitudes, the following parametrization for
the longitudinal and transverse polarization vectors is useful:
\begin{eqnarray}
\epsilon_2^L&=&\frac{1}{\sqrt{2(1-r_v^2)}r}(1-r_v^2,-r^2,\textbf{0}_{\rm T}),\quad \epsilon_2^T=(0,0,\textbf{1}_{\rm T}),\nonumber\\
\epsilon_3^L&=&\frac{1}{\sqrt{2(1-r^2)}r_v}(-r_v^2,1-r^2,\textbf{0}_{\rm T}),\quad \epsilon_3^T=(0,0,\textbf{1}_{\rm T}),
\end{eqnarray}
which satisfy the normalization $(\epsilon_{2,3}^{L})^2= (\epsilon_{2,3}^{T})^2=-1$  and the orthogonality $\epsilon_2^L \cdot P_2=\epsilon_3^L \cdot P_3=0$.

The decay amplitude can be decomposed into three parts of the polarizations amplitudes  as follows:
\begin{eqnarray}
\mathcal{A}(B\rightarrow \psi(2S) V)=\mathcal{A}_L+\mathcal{A}_N\epsilon_2^T\cdot\epsilon_3^T
+i\mathcal{A}_T\epsilon_{\alpha\beta\rho\sigma}n^{\alpha}v^{\beta}\epsilon_2^{T\rho}\epsilon_3^{T\sigma},
\end{eqnarray}
with the null vectors $n=(1,0,\textbf{0}_{\rm T})$ and $v=(0,1,\textbf{0}_{\rm T})$.
The subscript $L,N,T$  correspond to the  longitudinal, normal and transverse polarization states, respectively.
According to Eq.~(\ref{eq:operator}), the three different polarization amplitudes  have the following expressions,
\begin{eqnarray}\label{eq:alnt}
\mathcal{A}_{L,N,T}(B\rightarrow \psi(2S) V)&=&\xi_c
\Big [(C_1+\frac{1}{3}C_2)\mathcal{F}_{L,N,T}^{LL}+C_2\mathcal{M}_{L,N,T}^{LL} \Big]
-\xi_t\Big [(C_3+\frac{1}{3}C_4+C_9+\frac{1}{3}C_{10})\mathcal{F}_{L,N,T}^{LL}+\nonumber\\
&&(C_5+\frac{1}{3}C_6+C_7+\frac{1}{3}C_{8})\mathcal{F}_{L,N,T}^{LR}
+(C_4+C_{10})\mathcal{M}_{L,N,T}^{LL}+(C_6+C_8)\mathcal{M}_{L,N,T}^{SP}\Big ],
\end{eqnarray}
where $\mathcal{F}(\mathcal{M})$ describes the contributions from the factorizable (nonfactorizable ) diagrams.
The superscript $LL$, $LR$, and $SP$  refers to the contributions from $(V-A)\otimes(V-A)$, $(V-A)\otimes(V+A)$
and $(S-P)\otimes(S+P)$ operators, respectively. These explicit factorization formulas are all listed in Appendix A.
In this work, we also consider the vertex corrections to the factorizable amplitudes $\mathcal{F}$ at the current
known next-to-leading order (NLO) level. Their effects can be combined in the Wilson coefficients as usual \cite{bbns}.
In the NDR scheme,
the vertex corrections are included by the modifications to the combinations $a_i$
\footnote{The definitions of $a_2,a_{3,5,7,9}$ are of the form: $a_2=C_1+C_2/3$,
$a_i=C_i+C_{i+1}/3$ for $i=(3,5,7,9)$. } of the Wilson coefficients $C_i$
associated with the factorizable amplitudes in Eq.~(\ref{eq:alnt}):
\begin{eqnarray}\label{eq:vertex}
a_2&\rightarrow & a_2 +\frac{\alpha_s}{9\pi}C_2 \left[-18-12\text{ln}(\frac{\mu}{m_b})+f^h_I\right ],\nonumber\\
a_3+a_9 &\rightarrow &a_3+a_9+\frac{\alpha_s}{9\pi}(C_4+C_{10})\left [-18-12\text{ln}(\frac{\mu}{m_b})+f^h_I \right ],\nonumber\\
a_5+a_7&\rightarrow&a_5+a_7+\frac{\alpha_s}{9\pi}( C_6+C_{8})\left [6+12\text{ln}(\frac{\mu}{m_b})-f^h_I\right ].
\end{eqnarray}
The functions $f^h_I$  arise from the vertex corrections, which are given  in Ref.~\cite{prd65094023}.

\section{Numerical results}\label{sec:results}

To be used in our numerical calculations, those parameters such as meson mass,
the Wolfenstein parameters, decay constants, and the lifetime
of $B_{(s)}$ mesons \cite{pdg} are given in Table \ref{tab:constant1},
while the input wave functions and various parameters of the light vectors are shown in
Appendix \ref{sec:wave}.
\begin{table}
\caption{The decay constants of $\psi(2S)$ meson is from \cite{zhourui1}, while other parameters are adopted in PDG \cite{pdg}
in  our numerical calculations.  }
\label{tab:constant1}
\begin{tabular*}{14.5cm}{@{\extracolsep{\fill}}l|ccccc}
  \hline\hline
\textbf{Mass(\text{GeV})} & $M_{W}=80.385$   & $M_{B}=5.28$ & $M_{B_s}=5.37$ & $m_{b}=4.66$& $m_{c}=1.275$ \\[1ex]
& $m_{\psi(2S)}=3.686$&$m_{\rho}=0.775$&$m_{\omega}=0.783$&$m_{K^*}=0.892$&$m_{\phi}=1.019$     \\[1ex]
\hline
\end{tabular*}
\begin{tabular*}{14.5cm}{@{\extracolsep{\fill}}l|cccc}
  \hline
\multirow{2}{*}{{\textbf{The
Wolfenstein parameters}}}&$\lambda = 0.22506$,\quad &$A=0.811$,\quad &$\bar{\rho}=0.124$,\quad &$\bar{\eta}=0.356$ \\[1ex]
\hline
\end{tabular*}
\begin{tabular*}{14.5cm}{@{\extracolsep{\fill}}l|ccc}
\hline
\textbf{Decay constants(MeV)} & $f_{B}=190.9\pm 4.1$& $f_{B_s}=227.2\pm 3.4$ & $f_{\psi(2S)}=296^{+3}_{-2}$\\[1ex]
\hline
\end{tabular*}
\begin{tabular*}{14.5cm}{@{\extracolsep{\fill}}l|ccc}
\hline
\textbf{Lifetime(ps)} & $\tau_{B_s}=1.51$& $\tau_{B_0}=1.52$& $\tau_{B^+}=1.638$\\[1ex]
\hline\hline
\end{tabular*}
\end{table}
We now use the method previously illustrated to estimate
the physical observables (such as the $CP$ averaged branching ratios, direct and mixing $CP$ violations,
polarization fractions, and relative phases ) of the considered  decays.

\subsection{The $CP$ averaged branching ratios  }

For $B\rightarrow \psi(2S) V$ decays, the branching ratios   can be written as
\begin{eqnarray}
\mathcal {B}(B\rightarrow \psi(2S) V)=\frac{G_F^2\tau_{B}}{32\pi M}(1-r^2)\sum_{i=0,\parallel,\perp}|\mathcal {A}_i|^2,
\end{eqnarray}
where the terms $\mathcal {A}_0, \mathcal {A}_{\parallel},\mathcal {A}_{\perp}$ denote the longitudinal,
parallel, and perpendicular polarization amplitude in the trasversity
basis, respectively, which are related to $\mathcal {A}_{L,N,T}$ of Eq.~(\ref{eq:alnt}) via
\begin{eqnarray}\label{eq:spin}
\mathcal {A}_0=-\mathcal {A}_L, \quad \mathcal {A}_{\parallel}=\sqrt{2}\mathcal {A}_{N}, \quad \mathcal {A}_{\perp}=\sqrt{2}\mathcal {A}_{T}.
\end{eqnarray}
Here $\mathcal {A}_0$  and $ \mathcal {A}_{\parallel}$ are the $CP$ even amplitudes whereas $\mathcal {A}_{\perp}$
correspond to $CP$ odd ones.
Note that an additional minus sign in $\mathcal {A}_0$ (see Ref.~\cite{prd76074018} ) make our definitions of
the relative phase between $ \mathcal {A}_{\parallel(\perp)}$ and $ \mathcal {A}_{0}$ takes the value of $\pi$
in the heavy-quark limit.
The $CP$ averaged branching ratios for  the $B\rightarrow \psi(2S) V$ decays are shown in
Table \ref{tab:br1} together with some of the experimental measurements.
Some dominant uncertainties are considered in our calculations.
The first error in these entries is caused by the hadronic
parameters in the $B_{(s)}$ meson wave function: 
(1) the shape parameters: $\omega_b=0.40\pm 0.04$ for the $B$ meson, and $\omega_b=0.50\pm 0.05$ for the $B_s$
meson; (2) the decay constants, which are given in Table \ref{tab:constant1}.
The second error is from the uncertainty of the heavy quark masses.
In the evaluation, we vary the values of $m_{c(b)}$  within a $10\%$ range.
The last one is caused by the variation of the hard scale from
$0.8 t$ to $1.2 t$, which characterizes the size of the NLO QCD contributions.
It is found that the main uncertainties in our approach come from the $B$ meson wave function, which can reach $20-30\%$
in magnitude. The scale-dependent uncertainty is less than $20\%$ due
to the inclusion of the NLO vertex corrections.
We have checked the sensitivity of our results to the choice of the shape parameter
$\omega_c$ (see Eq.~(\ref{eq:wc})) in  charmonia meson wave function.
The variation of $\omega_c$ in the range $0.18\sim 0.22$ will result in a small change of the branching ratio,
say less than $10\%$.
In addition, the  uncertainties  related to the light vector mesons, such as the vector meson decay
constants and the Gegenbauer moments shown in Table \ref{tab:dc}, are only several percent.
Therefore they have been  neglected  in our calculations.

\begin{table}
\caption{
The PQCD predictions for the CP-averaged branching ratios for the  $B\rightarrow \psi(2S) V$ decays
( in units of $10^{-5}$). For comparison, experimental results from  BaBar  \cite{prl94141801},
Belle \cite{0308039,prd88074026}, or the world average from HFAG 2016 \cite{hfag}
and  PDG 2016 \cite{pdg} are also listed. }
\label{tab:br1}
\begin{tabular}[t]{l cccccc}
\hline\hline
Modes     & This work  & BaBar \cite{prl94141801} & Belle \cite{0308039} & Belle \cite{prd88074026} & HFAG 2016 \cite{hfag} & PDG 2016 \cite{pdg}
 \\ \hline
$B_s^0 \rightarrow \psi(2S)\bar{K}^{*0}$ &$2.2^{+0.6+0.2+0.3}_{-0.5-0.2-0.2}$   &-- &-- &--  &--    & $3.3\pm 0.5 $\\
$B_s^0 \rightarrow \psi(2S)\phi$  &$47^{+15+7+8}_{-10-3-4}$ &--&--&--  &-- & $54\pm 5 $\\
$B^+ \rightarrow \psi(2S)K^{*+}$ &$59^{+14+7+7}_{-12-7-5}$& $59.2\pm12.3$ & $81.3\pm 11.8$ &--  &$70.7\pm8.5$  & $67\pm1.4$\\
$B^0 \rightarrow \psi(2S)K^{*0}$   &$54^{+13+6+7}_{-11-6-5}$ &$64.9\pm 11.4$  &$72\pm 7.8$
&$55.5^{+4.7}_{-8.7}$ &$71.1\pm6.2$ &$59\pm4$\\ \hline
$B^+ \rightarrow \psi(2S)\rho^{+}$  &$2.7^{+0.6+0.3+0.3}_{-0.6-0.3-0.2}$ & -- & --& --&--  &-- \\
$B^0 \rightarrow \psi(2S)\rho^0$   &$1.2^{+0.3+0.1+0.1}_{-0.3-0.1-0.1}$ &--& --& -- &--  &-- \\
$B^0 \rightarrow \psi(2S)\omega$   &$1.0^{+0.2+0.1+0.1}_{-0.2-0.1-0.1}$ &-- & --& --&--  &-- \\
\hline\hline
\end{tabular}
\end{table}

For the  color-suppressed decays, it is expected
that the factorizable diagram contribution is suppressed due to the cancellation of Wilson coefficients $C_1+C_2/3$.
After the inclusion of the vertex corrections, the
factorizable diagram contributions become comparable with the nonfactorizable ones.
Some important features of the numerical results collected in Table \ref{tab:br1} are of the form
\begin{itemize}
\item[(I)]
The $b\rightarrow s$ transition processes $B^{+(0)}\rightarrow \psi(2S) K^{*+(0)}$ and
$B_s\rightarrow \psi(2S) \phi$  have a comparatively large branching ratio $10^{-4}$;
while the branching ratios of those $b\rightarrow d$ channels $B^{+(0)}\rightarrow\psi(2S)\rho^{+(0)}$,
$B^{0}\rightarrow\psi(2S)\omega^{0}$ and
$B_s\rightarrow \psi(2S) \bar{K}^{*0}$ are relatively small ( $\sim 10^{-5}$ ) owing to the
CKM factor suppression:  $|V^*_{cb}V_{cd}|\sim \lambda^3$.

\item[(II)]
In the quark model, the difference between $B^0 \rightarrow \psi(2S)\omega$  and $B^0 \rightarrow \psi(2S)\rho^0$
decays comes from the sign of $d\bar{d}$ component, which only appears in penguin operators,
so their difference should be relatively small.
The branching ratio ${\cal B}(B^0 \rightarrow \psi(2S)\omega)$  is indeed slightly smaller than
${\cal B}(B^0 \rightarrow \psi(2S)\rho^0)$.  This is a consequence of the fact
that the $\omega$ vector and tensor decay constants are smaller than those of the $\rho^0$ according to
Table \ref{tab:dc};

\item[(III)]
The value of $\mathcal {B}(B_s \rightarrow \psi(2S) \bar{K}^{*0} )$  have a tendency to be
smaller than 2$\mathcal {B}(B^0 \rightarrow \psi(2S) \rho^{0} )$.
Although  the $K^*$ and $B_s$ meson decay constants  are larger than those of the $\rho^0$ and $B^0$ meson,
the SU(3) breaking effects in the twist-2 distribution amplitudes, parametrized by the first Gegenbauer moment
$a_{1K^*}$ (see Eq. (\ref{eq:twist2})), gives a negative contribution to the $B_s \rightarrow \psi(2S) \bar{K}^{*0}$ decay,
which  induces the smaller  branching ratio.

\item[(IV)]
For the first four $B_{(s)}\to \psi(2S) V$ decays as listed in Table  \ref{tab:br1},
one can see that the PQCD predictions for their branching ratios agree well with the world averaged values
as given in HFAG 2016 and PDG 2016 \cite{hfag,pdg} within one standard deviation.
For $B_s \rightarrow \psi(2S) \bar{K}^{*0} $ decay, the central value of our theoretical prediction for
its branching ratio is slightly smaller than that of the PDG number \cite{pdg}.
But we know that the PDG result is obtained by multiplying the best value
$\mathcal {B}(B^{0}\rightarrow \psi(2S) K^{*0})$ with
the measured ratio $\mathcal {B}(\bar{B}_s^{0}\rightarrow \psi(2S) K^{*0})/\mathcal {B}(B^{0}\rightarrow
\psi(2S) K^{*0})$ from the LHCb \cite{plb747484}.
We hope the future experiment  will provide a direct measurement to this mode.

\item[(V)]
As for the channels with $\rho$ and $\omega$  as the final state $V$ meson, they have not been measured yet.
The pQCD predictions for the decay rates of these three channels are at the order of $10^{-5}$, measurable
in the future LHCb and Belle-II experiments.
\end{itemize}

For a more direct comparison with the available experimental measurements of the relative rates of
$B_{(s)}$ meson decays into $\psi(2S)$ and $J/\psi$ mesons, we recalculated the corresponding
$B_{(s)}$ decays to $J/\psi V$
by using the same  input parameters as in this paper but with  the replacement $\psi(2S)\rightarrow J/\psi$,
and we found numerically that
\begin{eqnarray}\label{eq:1s}
\mathcal {B}(B_s \rightarrow J/\psi \bar{K}^{*0} )&=&(4.2^{+1.2+0.6+0.6}_{-0.8-0.3-0.1})\times 10^{-5}, \non
\mathcal {B}(B_s \rightarrow J/\psi \phi )&=&(9.3^{+2.6+1.0+1.5}_{-1.9-0.7-0.8})\times 10^{-4},\non
\mathcal {B}(B^+ \rightarrow J/\psi K^{*+} )&=&(11.2^{+2.5+1.4+1.5}_{-2.2-1.2-0.9})\times 10^{-4}, \non
\mathcal {B}(B^0 \rightarrow J/\psi K^{*0} )&=&(10.4^{+2.2+1.3+1.3}_{-2.0-1.1-0.8})\times 10^{-4},\non
\mathcal {B}(B^+ \rightarrow J/\psi \rho^+ )&=&(5.1^{+1.2+0.6+0.8}_{-1.0-0.5-0.3})\times 10^{-5},\non
\mathcal {B}(B^0 \rightarrow J/\psi \rho^0 )&=&(2.4^{+0.6+0.3+0.4}_{-0.5-0.3-0.2})\times 10^{-5}, \non
\mathcal {B}(B^0 \rightarrow J/\psi \omega )&=&(1.8^{+0.4+0.2+0.3}_{-0.4-0.1-0.1})\times 10^{-5},
\end{eqnarray}
where the errors have the same meaning as those for $B_{(s)}\to \psi(2S) V$ decays.
The above results are well consistent with the previous PQCD calculations \cite{prd89094010} and also
the present data \cite{pdg}.

Finally, as a cross-check, using the PQCD predictions as given in Table \ref{tab:br1} and Eq.~(\ref{eq:1s})
we can estimate the relative ratios
$\mathcal {R}_V=\mathcal {B}(B\rightarrow \psi(2S)V)/\mathcal {B}(B\rightarrow J/\psi V)$ as below,
 \begin{eqnarray}
\mathcal {R}_{\phi}&=&0.51^{+0.02}_{-0.01}, \quad \mathcal {R}_{K^{*0(+)}}=0.53^{+0.00}_{-0.02}, \quad
\mathcal {R}_{\rho^{0(+)}}=0.53^{+0.00}_{-0.03},\non
\mathcal {R}_{\omega}&=&0.56^{+0.01}_{-0.04}, \quad \mathcal {R}_{\bar{K}^{*0}}=0.52^{+0.01}_{-0.04},
\end{eqnarray}
where all uncertainties are added in quadrature.
Since the parameter dependences of the PQCD predictions for the branching ratios  are largely
canceled in their relative ratios, the total theoretical error of $R_{V}$ are only a few percent,
much smaller than those for the branching ratios.
Fortunately, two of these five ratios have been measured by LHCb \cite{epjc722118} D0 \cite{prd79111102}, and CDF \cite{prd58072001,prl96231801} experiments:
\begin{eqnarray}
\mathcal {R}_{\phi}=\left\{
\begin{aligned}
0.489& \pm 0.026(\text{stat})\pm 0.021(\text{syst})\pm0.012(R_{\psi})\quad\quad\quad  &\text{LHCb} \nonumber\\ 
0.53&\pm0.10(\text{stat})\pm0.07(\text{syst})\pm0.06(R_{\psi})\quad\quad\quad\quad &\text{D0 } \\ 
0.52&\pm0.13(\text{stat})\pm0.04(\text{syst})\pm0.06(R_{\psi})\quad\quad\quad\quad &\text{ CDF }
\end{aligned}\right.
\end{eqnarray}
 \begin{eqnarray}
\mathcal {R}_{K^{*0}}=\left\{
\begin{aligned}
0.476&\pm0.014(\text{stat})\pm0.010(\text{syst})\pm0.012(R_{\psi})\quad\quad\quad &\text{LHCb} \\
0.515&\pm0.113(\text{stat})\pm0.052(\text{syst}) \quad\quad\quad\quad &\text{ CDF }, 
\end{aligned}\right.
\end{eqnarray}
where the third uncertainty is from the ratio of the $\psi(2S)$
and $J/\psi$ branching fractions to $\mu^+\mu^-$.
It is easy to see that our PQCD predictions for both $R_\phi$ and $R_{K^{*0}}$ agree very well with the
measured values.

\subsection{CP ASYMMETRIES }

Studying $CP$ asymmetries is an important task in $B$ physics. For the charged  $B$ decays, the $CP$ asymmetries arise from the interference between the
penguin diagrams and tree diagrams.
The direct $CP$ violation asymmetry including three polarization  are  defined by
 \begin{eqnarray}
A^{\text{dir}}_{0,\parallel,\perp}=\frac{|\mathcal {\bar{A}}_{0,\parallel,\perp}|^2-|\mathcal {A}_{0,\parallel,\perp}|^2}{|\mathcal {\bar{A}}_{0,\parallel,\perp}|^2+|\mathcal {A}_{0,\parallel,\perp}|^2},
\end{eqnarray}
where $\mathcal {\bar{A}}_{0,\parallel,\perp}$ is the $CP$-conjugate amplitude of $\mathcal {A}_{0,\parallel,\perp}$.

For the neutral $B^0_{(s)}$ decays, because of the $B^0_{(s)}-\bar{B}^0_{(s)}$ mixing, it
is required to include time-dependent measurements in
$CP$  violation asymmetries. If the final states are $CP$ eigen
states, the time-dependent $CP$ asymmetry is defined as
 \begin{eqnarray}
A^{0,\parallel,\perp}_f(t)=-C^{0,\parallel,\perp}_f \text{cos}(\Delta m t)+S^{0,\parallel,\perp}_f \text{sin}(\Delta m t),
\end{eqnarray}
where $\Delta m$ is the mass difference of the two mass eigenstates
of the neutral $B$ meson and   $f$ is a two-body final state.
The direct $CP$ asymmetry  $C^{0,\parallel,\perp}_f$
and mixing-induced $CP$ asymmetry $S^{0,\parallel,\perp}_f$ are referred to as
 \begin{eqnarray}\label{eq:violation}
C^{0,\parallel,\perp}_f=\frac{1-|\lambda^{0,\parallel,\perp}_f|^2}{1+|\lambda^{0,\parallel,\perp}_f|^2},\quad S_f=\frac{2\text{Im}(\lambda^{0,\parallel,\perp}_f)}{1+|\lambda^{0,\parallel,\perp}_f|^2}.
\end{eqnarray}
The parameter $\lambda^{0,\parallel,\perp}_f=\eta_f e^{-2i\beta_{(s)}}\frac{\mathcal {\bar{A}}_{0,\parallel,\perp}}{\mathcal {A}_{0,\parallel,\perp}}$ describes $CP$ violation in the interference between mixing and decay.
$\eta_f$ is the $CP$ eigenvalue $(\pm1)$ of the polarization state. $\beta_{(s)}$ is the CKM angle defined as usual \cite{pdg}.
Note that  the final states of $\psi(2S) K^{*0}$ and its $CP$ conjugate are flavor-specific, for example, the kaon and pion
charges of $K^{*0}\rightarrow K^+\pi^-$ and $\bar{K}^{*0}\rightarrow K^-\pi^+$
depend on whether we had a $B$ and $\bar{B}$ 
meson in the initial state,  and the time-dependent angular distributions do not show $CP$ violation due to interference between mixing and decay.
Therefore, we only calculate the direct $CP$ asymmetry for $B^0_s\rightarrow \psi(2S) \bar{K}^{*0}$ and $B^0\rightarrow \psi(2S) K^{*0}$ decays.

The PQCD predictions for the $CP$ asymmetry parameters $A^{\text{dir}}_{0,\parallel,\perp}$ are listed in
Table \ref{tab:dir} and  \ref{tab:mix}.
Unlike the branching ratios, the direct $CP$ asymmetry is not sensitive to the wave function parameters
and heavy quark masses, but suffer from large uncertainties due to the hard scale $t$.
In order to reduce the large scale dependence effectively, one has to know the complete NLO  corrections,
which are in fact not yet available now, thus beyond the scope of this paper.

Since the direct $CP$ asymmetry is proportional to the interference between the tree and penguin contributions,
while the Wilson coefficients of the penguin diagram are loop suppressed when compared with those tree contributions.
Therefore, the direct $CP$ asymmetry parameters of  these processes are rather small ( only $10^{-3}\sim 10^{-4}$),
and  the mixing-induced $CP$ asymmetry for neutral $B$ decays is almost proportional to the $\sin2\beta_{(s)}$
from Eq. (\ref{eq:violation}).
The  mixing-induced $CP$ asymmetry parameters $S_{\psi(2S)\rho^0(\omega)}$ and $S_{\psi(2S)\phi}$ in Table \ref{tab:mix}
are very close to  the current world average values $-\sin{2\beta}=-0.691\pm 0.017$ and $-2\beta_s=-0.0376^{+0.0008}_{-0.0007}$ \cite{hfag}, respectively.
That is to say, these modes can serve as  an alternative places to extract CKM angle $\beta_{(s)}$.
Furthermore, the large  mixing-induced $CP$ asymmetry $S_f$ for $b\rightarrow d$ transition can confront with future experimental results.
It can also be seen that the  $CP$  asymmetry parameters for three polarization states are slightly different
because the strong phases coming from the non-factorizable diagrams and vertex corrections
are polarization-dependent \cite{prd65094023}.
On experimental side, so far  only the  charge asymmetries of $B^+\rightarrow \psi(2S) K^{*+}$ process
was measured by BaBar Collaboration \cite{pdg}:
\begin{eqnarray}
A^{\text{dir}}_{CP}(B^+\rightarrow \psi(2S) K^{*+})=0.08\pm 0.21.
\end{eqnarray}
Of course, the statistical uncertainty is too large  to make any statement.
Any observation of large direct $CP$ asymmetry for the considered decays $B_{(s)}\to \psi(2S) V$ decays
will be a signal for new physics.
Besides, the  precise measurements of these mixing-induced $CP$
asymmetries serve to determine the $CP$ phases related to the $B^0-\bar{B^0}$ and $B^0_s-\bar{B^0_s}$ mixing amplitudes.

\begin{table}
\caption{The PQCD predictions for $A^{\text{dir}}_{0,\parallel,\perp}(10^{-3})$
in the $B\to \psi(2S) (\rho^+,K^*, \bar{K}^* )$ decays. The error arises from the hard scale $t$.}
\label{tab:dir}
\begin{tabular}[t]{l cccc}
\hline\hline
Modes     &$A^{\text{dir}}_{0}$ &$A^{\text{dir}}_{\parallel}$&$A^{\text{dir}}_{\perp}$ &$A^{\text{dir}}$
 \\ \hline
$B^+ \rightarrow \psi(2S)\rho^{+}$  &$-5.9^{+6.6}_{-11.9}$ & $-8.3^{+6.9}_{-7.0}$ & $-9.2^{+5.7}_{-11.3}$& $-7.2^{+6.6}_{-10.5}$\\
$B^+ \rightarrow \psi(2S)K^{*+}$  &$0.4^{+0.6}_{-0.4}$ & $0.4^{+1.3}_{-0.4}$ & $0.5^{+0.3}_{-0.3}$& $0.4^{+0.8}_{-0.4}$\\
$B^0_s \rightarrow \psi(2S)\bar{K}^{*0}$  &$-5.2^{+7.2}_{-9.2}$ & $-5.7^{+6.7}_{-7.0}$ & $-7.1^{+5.7}_{-9.4}$& $-5.7^{+5.8}_{-8.7}$\\
$B^0 \rightarrow \psi(2S)K^{*0}$  &$0.4^{+0.6}_{-0.4}$ & $0.4^{+1.3}_{-0.4}$ & $0.5^{+0.3}_{-0.3}$& $0.4^{+0.8}_{-0.4}$\\
\hline\hline
\end{tabular}
\end{table}

\begin{table}
\caption{The PQCD predictions for the $CP$ asymmetry parameters  $C_f^{0,\parallel,\perp}$ and $S_f^{0,\parallel,\perp}$
in the $B^0\rightarrow \psi(2S) ( \rho^0, \omega, \phi)$ decays. The error arises from the hard scale $t$.}
\label{tab:mix}
\begin{tabular}[t]{l cccccc}
\hline\hline
Modes     &$C_f^{0}(10^{-3})$ &$S_f^{0}$ &$C_f^{\parallel}(10^{-3})$ &$S_f^{\parallel}$ &$C_f^{\perp}(10^{-3})$ &$S_f^{\perp}$
 \\ \hline
$B^0 \rightarrow \psi(2S)\rho^{0}$  &$5.9^{+11.2}_{-6.7}$& $-0.68^{+0.00}_{-0.01}$ &$8.3^{+6.6}_{-5.9}$
&$-0.69^{+0.00}_{-0.00}$ &$9.2^{+11.8}_{-6.1}$&$0.69^{+0.01}_{-0.00}$  \\
$B^0 \rightarrow \psi(2S)\omega$  &$6.4^{+7.7}_{-6.0}$& $-0.68^{+0.00}_{-0.01}$ &$7.9^{+8.9}_{-6.7}$ &$-0.69^{+0.00}_{-0.01}$
&$8.9^{+10.2}_{-5.7}$&$0.69^{+0.01}_{-0.00}$  \\
$B^0_s \rightarrow \psi(2S)\phi$  &$-0.4^{+0.3}_{-0.4}$& $-0.038^{+0.001}_{-0.000}$
&$-0.3^{+0.2}_{-0.5}$ &$-0.038^{+0.001}_{-0.000}$ &$-0.4^{+0.3}_{-0.3}$&$0.037^{+0.000}_{-0.001}$  \\
\hline\hline
\end{tabular}
\end{table}

\subsection{Polarization fractions and  relative phases}
\begin{table}
\caption{The PQCD predictions for the CP-averaged polarization fractions, relative phases in the
$B\rightarrow \psi(2S) V$  decays.
The errors correspond to the combined uncertainty in the hadronic parameters, heavy quark masses and the hard scale $t$.}
\label{tab:po1}
\begin{tabular}[t]{l cccccc}
\hline\hline
Modes   & $f_0$ & $f_{\parallel}$ &$ f_{\perp}$ &$ \phi_{\parallel}(\text{rad})$&$\phi_{\perp}(\text{rad})$
 \\ \hline
 $B^{+} \rightarrow \psi(2S)K^{*+}$  & $0.48^{+0.01+0.07+0.01}_{-0.02-0.08-0.01}$ & $0.28^{+0.01+0.03+0.00}_{-0.00-0.03-0.00}$
  & $0.24^{+0.01+0.05+0.00}_{-0.00-0.05-0.00}$  &$2.43^{+0.01+0.09+0.03}_{-0.02-0.09-0.04}$ &$2.15^{+0.02+0.16+0.01}_{-0.03-0.16-0.05}$   \\
CLEO \cite{prd63031103}   &$0.51\pm0.16\pm0.05$&--&--&--&--\\ \hline
$B^{0} \rightarrow \psi(2S)K^{*0}$  & $0.48^{+0.01+0.07+0.01}_{-0.02-0.08-0.01}$ & $0.28^{+0.01+0.03+0.00}_{-0.00-0.03-0.00}$
  & $0.24^{+0.01+0.05+0.00}_{-0.00-0.05-0.00}$  &$2.43^{+0.01+0.09+0.03}_{-0.02-0.09-0.04}$ &$2.15^{+0.02+0.16+0.01}_{-0.03-0.16-0.05}$   \\
BaBar \cite{prd76031102}   &$0.48\pm0.05\pm0.02$&$0.22\pm0.06\pm0.02$&$0.30\pm0.06\pm0.02$&$3.5\pm0.4\pm0.1$ \footnotemark[1] &$2.8\pm0.3\pm0.1$\\
CLEO \cite{prd63031103}   &$0.40\pm0.14\pm0.07$&--&--&--&--\\
Belle \cite{prd88074026}   &$0.455^{+0.031+0.014}_{-0.029-0.049}$&--&--&--&--\\ \hline
$B^0_s \rightarrow \psi(2S)\bar{K}^{*0}$   & $0.50^{+0.01+0.06+0.01}_{-0.02-0.07-0.01}$ & $0.28^{+0.00+0.02+0.00}_{-0.00-0.04-0.01}$
 & $0.23^{+0.01+0.04+0.00}_{-0.01-0.04-0.01}$  &$2.48^{+0.01+0.08+0.02}_{-0.02-0.08-0.04}$&$2.20^{+0.03+0.16+0.05}_{-0.02-0.13-0.05}$   \\
LHCb \cite{plb747484} &$0.524\pm0.056\pm0.029$& --&--&--&--\\ \hline
$B^0_s \rightarrow \psi(2S)\phi$   & $0.48^{+0.01+0.05+0.00}_{-0.02-0.06-0.01}$ & $0.29^{+0.00+0.02+0.00}_{-0.01-0.03-0.01}$
& $0.24^{+0.00+0.04+0.00}_{-0.01-0.04-0.01}$  &$2.59^{+0.01+0.08+0.02}_{-0.01-0.05-0.03}$ &$2.31^{+0.02+0.14+0.03}_{-0.02-0.11-0.04}$   \\
LHCb \cite{plb762253} &$0.422\pm 0.014\pm 0.003$ & --&$0.264^{+0.024}_{-0.023}\pm 0.002 $&$3.67^{+0.13}_{-0.18}
\pm 0.03$&$3.29^{+0.43}_{-0.39}\pm0.04$\\ \hline
$B^{+} \rightarrow \psi(2S)\rho^{+} $  & $0.54^{+0.01+0.06+0.01}_{-0.02-0.08-0.00}$ & $0.25^{+0.01+0.03+0.01}_{-0.01-0.03-0.01}$
& $0.21^{+0.01+0.05+0.00}_{-0.01-0.04-0.00}$  &$2.32^{+0.02+0.12+0.03}_{-0.02-0.12-0.03}$&$2.05^{+0.02+0.17+0.04}_{-0.04-0.20-0.06}$   \\
$B^{0} \rightarrow \psi(2S)\rho^{0} $ & $0.54^{+0.01+0.06+0.01}_{-0.02-0.08-0.00}$ & $0.25^{+0.01+0.03+0.01}_{-0.01-0.03-0.01}$
& $0.21^{+0.01+0.05+0.00}_{-0.01-0.04-0.00}$  &$2.32^{+0.02+0.12+0.03}_{-0.02-0.12-0.03}$&$2.05^{+0.02+0.17+0.04}_{-0.04-0.20-0.06}$   \\
$B^{0} \rightarrow \psi(2S)\omega^{0} $  & $0.52^{+0.02+0.08+0.01}_{-0.01-0.07-0.00}$ & $0.25^{+0.01+0.03+0.01}_{-0.00-0.02-0.00}$
& $0.22^{+0.01+0.05+0.00}_{-0.01-0.04-0.01}$ &$2.34^{+0.02+0.11+0.05}_{-0.02-0.12-0.03}$&$2.07^{+0.02+0.17+0.04}_{-0.03-0.19-0.06}$   \\
\hline\hline
\end{tabular}
\footnotetext[1]{We choose the equivalent solution in $(0,2\pi)$}
\end{table}

In experimental analyses, we usually define five observables corresponding to three polarization fractions $f_0, f_{\parallel}, f_{\perp}$,
and two relative phases $\phi_{\parallel}, \phi_{\perp}$, where
 \begin{eqnarray}\label{eq:jihua}
f_{0,\parallel,\perp}=\frac{|\mathcal {A}_{0,\parallel,\perp}|^2}{|\mathcal {A}_0|^2+|\mathcal {A}_{\parallel}|^2+|\mathcal {A}_{\perp}|^2},\quad
\phi_{\parallel,\perp}=\text{arg}\frac{\mathcal {A}_{\parallel,\perp}}{\mathcal {A}_{0}},
\end{eqnarray}
with normalisation such that $f_0+f_{\parallel}+f_{\perp}=1$.
The polarization fractions as well as relative phases are shown in Table \ref{tab:po1}, where the sources of the errors in the
 numerical estimates have the same origin as in the discussion of the branching ratios in Table \ref{tab:br1}.
 It is easy to see that the most important theoretical
uncertainties are caused by the heavy quark masses. From  Eqs. (\ref{eq:vertex}) and (\ref{eq:mm}), we can see the mass terms $m_b$ and $m_c$ 
associated vertex corrections  and nonfactorizable amplitudes, respectively. It can numerically change the real and imaginary parts of these contributions and
have a significant effect on the polarization fractions, especially for the relative phases.
The uncertainties from the wave function parameters are very small because they  mainly give an overall change of all polarization
amplitudes  and   the parameter dependence can be canceled out in Eq. (\ref{eq:jihua}).

From Table \ref{tab:po1}, both the $B^+\rightarrow \psi(2S) (K^{*+},\rho^+)$  and  $B^0\rightarrow \psi(2S) (K^{*0},
\rho^0)$ modes have the same polarization fractions and relative phases,
since they differ only in the lifetimes or isospin factor in our formalism.
Comparing the three polarization fractions,
the perpendicular polarization fractions  $f_{\perp}$ are less than $25\%$  shows that the $CP$ even
component dominates in these decays.
According to the power counting rules in the factorization assumption, the longitudinal polarization dominates the
decay ratios and the transverse polarizations are suppressed
\cite{zpc1269} due to the helicity flips of the quark in the final state hadrons.
However, the situation is very different for the color-suppressed decays, where
the contributions from the nonfactorizable tree diagrams  in Figs.~\ref{fig:femy}(c) and \ref{fig:femy}(d)
are comparable with those of the color-suppressed tree diagrams  although the latter are enhanced
by the involving vertex corrections.
With an additional gluon, the transverse polarization in the
nonfactorizable diagrams does not encounter helicity flip suppression,
therefore numerically we get a longitudinal polarization fraction ($f_0$) smaller than $50\%$,
which are compatible with those currently available data.
The fact that the nonfactorizable diagrams can give a large transverse polarization
contribution is also observed in the $B_c\rightarrow J/\psi D_{(s)}^{*+}$ decays \cite{prd90114030}.
There are another equivalent set of helicity amplitudes $(\mathcal {A}_{0},\mathcal {A}_{+},\mathcal {A}_{-})$, which
are related to the spin amplitudes $(\mathcal {A}_{0},\mathcal {A}_{\parallel},\mathcal {A}_{\perp})$
introduced in Eq.~(\ref{eq:spin}) by
\begin{eqnarray}
\label{eq:helicity}
\mathcal {A}_{\pm}=\frac{\mathcal {A}_{\parallel}\pm\mathcal {A}_{\perp}}{\sqrt{2}},
\end{eqnarray}
while $\mathcal {A}_{0}$ is common to both bases.

It is expected that $|\mathcal {A}_{0}|^2 >|\mathcal {A}_{+}|^2>|\mathcal {A}_{-}|^2$
if the  two final-states are both  light vector mesons. 
The larger the mass of the vector-meson daughters, the weaker the inequality.
In $B\rightarrow \psi(2S) V$ decays with light $V$ being a recoiled meson and heavy $\psi(2S)$ an ejected one.
The positive-helicity amplitude is suppressed by $m_{\psi(2S)}/M$ (almost of order unity) due to one of  the quark helicities in $\psi(2S)$ has to be flipped,
while the negative-helicity one is subject to a further chirality suppression of order $m_V/M$ \cite{zpc1269}.
Therefore, $\mathcal {A}_{+}$ and $\mathcal {A}_{0}$ can be comparable and  larger than $\mathcal {A}_{-}$.
Using values of Table \ref{tab:po1} and Eq. (\ref{eq:helicity}),
the pQCD predictions do favor the hierarchy pattern $|\mathcal {A}_{0}|^2\sim |\mathcal {A}_{+}|^2>|\mathcal {A}_{-}|^2$.

The angular analysis of $B^0\rightarrow \psi(2S) K^{*0}$ and $B_s^0\rightarrow \psi(2S) \phi$
 has been carried out by BaBar \cite{prd76031102} and LHCb \cite{plb762253}, respectively. The
obtained polarization observables are also summarized in Table \ref{tab:po1}.
As expected under $SU(3)$-flavor symmetry, both decay modes have  similar magnitudes and phases of the amplitudes.
Our results of polarization fractions
can accommodate the data well within uncertainties,
 while the predicted relative phases are a bit smaller than the data.
One can find a shift from $\pi$ at the $6-7\sigma$ level
in $\phi_{\parallel}$ and  $\phi_{\perp}$ shows
the existence of  final-state interaction. However, the $f_{\parallel}-f_{\perp}$ is about $5\%$ and
the difference between $\phi_{\parallel}$ and $\phi_{\perp}$ dose not exceed $0.3$ radians,
which suggest  that our solutions are consistent with   approximate
s-quark helicity conservation despite substantial strong phases.

For the  $B^0_s \rightarrow \psi(2S)\bar{K}^{*0}$ channel,
the LHCb Collaboration \cite{plb747484}  has reported the longitudinal polarisation fraction $f_0$
as $0.524\pm0.056\pm0.029$, but a thorough angular analysis is still missing.
As for other  modes, we obtain reasonably accurate results, which could be tested by future experimental measurements.

\section{ conclusion}\label{sec:conlusion}

In this paper we have investigated the seven $B\rightarrow \psi(2S) V$ decay modes carefully by employing
the PQCD factorization approach.
Besides the color-suppressed factorizable diagrams, the nonfactorizable
diagrams and the vertex correction diagrams can also be evaluated in this approach.

The predicted branching ratios and the relative rates of  $B$ meson decays into $\psi(2S)$
and $J/\psi$ mesons  are compared with experiments wherever available.
Our results indicate that the direct $CP$ asymmetries in
these channels are very small due to  the suppressed penguin contributions as we mentioned above.
The mixing-induced $CP$ asymmetries are not far away from $\sin 2\beta_{(s)}$,
these channels can therefore play an important role in the extraction of the CKM angle $\beta_{(s)}$.

Finally, we made a comprehensive polarization analysis of  the considered decays.
The predicted  polarization fractions and  relative phases of $B^0\rightarrow \psi(2S) K^{*0}$ and
$B^0_s\rightarrow \psi(2S) \phi$  decays are consistent with data.
Due to the large mass of $\psi(2S)$ and the dominant contributions from the nonfactorizable diagrams,
we obtain an equal amount of transverse and longitudinal polarization.
The pattern of $f_{\parallel}\approx f_{\perp}$,  $ \phi_{\parallel}\approx \phi_{\perp}$ favor
the conservation of light quark helicity.
The deviations from $\pi$  at several standard deviations
in $\phi_{\parallel}$ and $\phi_{\perp}$ indicate the existence of the still unknown final-state interaction.

We also discussed theoretical uncertainties arising from
the hadronic parameters in $B$ meson wave function, heavy quark masses and hard scale $t$.
The total uncertainties are acceptable, around $30\%$ in magnitude.
The   uncertainties from the hadronic parameters can give sizable effects on the
PQCD predictions for branching ratios, while the $CP$ asymmetries suffer a large error from
the hard scale $t$. The further studies at the completely  NLO level are certainly required to
improve the accuracy of the theoretical predictions.
Furthermore, the polarization observables $f_{0,\parallel,\perp}$ and $\phi_{\parallel,\perp}$ are
more sensitive  to the heavy quark masses, which suggest that the color-suppressed type decays
may be more sensitive to the vertex corrections and  nonfactorizable contributions.
Our results and findings will be further tested by the LHCb  and Belle-II experiments in the near future.

\begin{acknowledgments}

The authors are grateful to Hsiang-nan Li
for helpful discussions. This work is supported in part by National Natural Science Foundation of China under
Grants Nos. 11547020, 11605060, and 11235005, in part by Natural Science Foundation of Hebei Province
under Grant No. A2014209308,
in part by  Program for the Top Young Innovative Talents of Higher Learning Institutions of Hebei
Educational Committee under Grant No. BJ2016041, and in part by Training Foundation of  North
China University of Science and Technology  under Grant No. GP201520 and No. JP201512.
\end{acknowledgments}

\begin{appendix}

\section{THE DECAY AMPLITUDES}\label{sec:ampulitude}

Following the derivation of the factorization formula of Eq. (\ref{eq:ampu}), 
 we get the analytic formulas of the (non)factorizable amplitude for each helicity state listed below.
\begin{eqnarray}\label{eq:flll}
\mathcal{F}^{LL}_L&=&-8\pi C_f f_{\psi} M^4\int_0^1dx_1dx_3\int_0^{\infty}b_1b_3db_1db_3 \phi_B(x_1,b_1)\nonumber\\&&\{
\sqrt{1-r^2}[\phi_V(x_3)((r^2-1)x_3-1)+\phi_V^s(x_3)\sqrt{1-r^2}r_v(2x_3-1)+\nonumber\\&&
\phi_V^t(x_3)r_v(2x_3-1-r^2(1+2x_3))]\alpha_s(t_a)S_{ab}(t_a)h(\alpha_e,\beta_a,b_1,b_3)S_t(x_1)\nonumber\\&&
-2r_{v}(1-r^2)\phi_V^s(x_3)\alpha_s(t_b)S_{ab}(t_b)h(\alpha_e,\beta_b,b_1,b_3)S_t(x_3)
\},
\end{eqnarray}
\begin{eqnarray}\label{eq:flln}
\mathcal{F}^{LL}_N&=&8\pi C_f f_{\psi} M^4r\int_0^1dx_1dx_3\int_0^{\infty}b_1b_3db_1db_3 \phi_B(x_1,b_1)\nonumber\\&&\{
[(r^2-1)(\phi_V^a(x_3)r_vx_3-\phi_V^T(x_3))+r_v\phi_V^v(2+(1-r^2)x_3)]\alpha_s(t_a)S_{ab}(t_a)h(\alpha_e,\beta_a,b_1,b_3)S_t(x_1)\nonumber\\&&
+r_v(1-r^2)(\phi_V^a(x_3)+\phi_V^v(x_3))\alpha_s(t_b)S_{ab}(t_b)h(\alpha_e,\beta_b,b_1,b_3)S_t(x_3)
\},
\end{eqnarray}
\begin{eqnarray}
\mathcal{F}^{LL}_T=\mathcal{F}^{LL}_N|_{\phi_V^a\leftrightarrow \phi_V^v},
\end{eqnarray}
\begin{eqnarray}
\mathcal{F}^{LR}_{L,N,T}=\mathcal{F}^{LL}_{L,N,T},
\end{eqnarray}
\begin{eqnarray}\label{eq:mlll}
\mathcal{M}^{LL}_L&=&-16\sqrt{\frac{2}{3}}\pi C_f M^4\int_0^1dx_1dx_2dx_3\int_0^{\infty}b_1b_2db_1db_2 \phi_B(x_1,b_1)\nonumber\\&&\{
\sqrt{1-r^2}[\psi^L(x_2,b_2)r_v(\phi_V^s(x_3)\sqrt{1-r^2}+\phi_V^t(x_3)(r^2(2x_2+x_3-2)-x_3))\nonumber\\&&
-\phi_V(x_3)(\psi^L(x_2,b_2)(r^2-1)(x_2-1)+\psi^t(x_2,b_2)r_cr]\alpha_s(t_c)S_{cd}(t_c)h(\alpha_e,\beta_c,b_1,b_2)\nonumber\\&&
+[\psi^L(x_2,b_2)(x_2(\phi_V(x_3)(r^2+1)-2r^2r_v\phi_V^t(x_3))-(r^2-1)x_3(\phi_V(x_3)-r_v\phi_V^t(x_3))\nonumber\\&&
-\phi_V^s(x_3)\sqrt{1-r^2}r_vx_3)-\psi^t(x_2,b_2)r_cr(\phi_V(x_3)-4r_v\phi_V^t(x_3))]\alpha_s(t_d)S_{cd}(t_d)h(\alpha_e,\beta_d,b_1,b_2)
\},
\end{eqnarray}
\begin{eqnarray}\label{eq:mlln}
\mathcal{M}^{LL}_N&=&16\sqrt{\frac{2}{3}}\pi C_f M^4 \int_0^1dx_1dx_2dx_3\int_0^{\infty}b_1b_2db_1db_2 \phi_B(x_1,b_1)\nonumber\\&&\{
[(r^2-1)(r_cr_v\psi^T(x_2,b_2)\phi_V^a(x_3)+r(x_2-1)\psi^V(x_2,b_2)\phi_V^T(x_3))\nonumber\\&&+\psi^T(x_2,b_2)
\phi_V^v(x_3)r_cr_v(1+r^2)]\alpha_s(t_c)S_{cd}(t_c)h(\alpha_e,\beta_c,b_1,b_2)\nonumber\\&&
+[\phi_V^v(x_3)r_v(\psi^T(x_2,b_2)r_c(1+r^2)-2\psi^V(x_2,b_2)r(x_2(1+r^2)+x_3(1-r^2)))\nonumber\\&&
-(r^2-1)(\phi_V^T(x_3)(\psi^V(x_2,b_2)rx_2-2\psi^T(x_2,b_2)r_c)+r_cr_v\phi_V^a(x_3)
\psi^T(x_2,b_2)]\nonumber\\&&\alpha_s(t_d)S_{cd}(t_d)h(\alpha_e,\beta_d,b_1,b_2)
\},
\end{eqnarray}
\begin{eqnarray}\label{eq:mllt}
\mathcal{M}^{LL}_T=\mathcal{M}^{LL}_N|_{\phi_V^a\leftrightarrow \phi_V^v},
\end{eqnarray}
\begin{eqnarray}\label{eq:msp}
\mathcal{M}^{SP}_{L,N,T}=-\mathcal{M}^{LL}_{L,N,T},
\end{eqnarray}
with $r_c=m_c/M$ and $m_c$ is the charm quark mass;  $C_f=4/3$ is a color factor; $f_{\psi}$ is the decay constant of the $\psi(2S)$ meson.
The coefficient $(-)\frac{1}{\sqrt{2}}$ appears for $B\rightarrow \psi(2S)(\rho^0) \omega$ decay, because only the  $d$ quark component of the
  $(\rho^0)\omega$ meson is involved.
We  neglect terms higher than $r^2_v$ orders, since the vector light cone wave functions derived from sum rules are expanded to this order \cite{LCDAs}.
 The functions $h$  come from the Fourier transform  of virtual quark and gluon
propagators. They are defined by
\begin{eqnarray}
h(\alpha,\beta,b_1,b_2)&=&h_1(\alpha,b_1)\times h_2(\beta,b_1,b_2),\nonumber\\
h_1(\alpha,b_1)&=&\left\{\begin{array}{ll}
K_0(\sqrt{\alpha}b_1), & \quad  \quad \alpha >0,\\
K_0(i\sqrt{-\alpha}b_1),& \quad  \quad \alpha<0,
\end{array} \right.\nonumber\\
h_2(\beta,b_1,b_2)&=&\left\{\begin{array}{ll}
\theta(b_1-b_2)I_0(\sqrt{\beta}b_2)K_0(\sqrt{\beta}b_1)+(b_1\leftrightarrow b_2), & \quad   \beta >0,\\
\theta(b_1-b_2)J_0(\sqrt{-\beta}b_2)K_0(i\sqrt{-\beta}b_1)+(b_1\leftrightarrow b_2),& \quad   \beta<0,
\end{array} \right.
\end{eqnarray}
where $J_0$ is the Bessel function and $K_0$, $I_0$ are modified Bessel function with
$K_0(ix)=\frac{\pi}{2}(-N_0(x)+i J_0(x))$.
$\alpha_e$ and $\beta_{a,b,c,d}$ are the virtuality of the internal gluon and quark, respectively. Their expressions are
\begin{eqnarray}\label{eq:mm}
\alpha_e &=&x_1x_3(1-r^2)M^2, \quad \beta_a=x_3(1-r^2)M^2, \quad \beta_b=x_1(1-r^2)M^2,
 \nonumber\\ \beta_c&=&[(x_1+x_2-1)(x_3+r^2(1-x_2-x_3))+r_c^2]M^2,
 \nonumber\\ \beta_d&=&[(x_1-x_2)(x_3+r^2(x_2-x_3))+r_c^2]M^2.
\end{eqnarray}
The hard scale $t$ is chosen as the maximum of the virtuality of the internal momentum transition in the hard amplitudes,
including $1/b_i(i=1,2,3)$:
\begin{eqnarray}
t_{a,b}&=&\max(\sqrt{\beta_{a,b}},1/b_1,1/b_3), \quad t_{c,d}=\max(\sqrt{\alpha_e},\sqrt{\beta_{c,d}},1/b_1,1/b_2).
\end{eqnarray}
The Sudakov factors can be written as
\begin{eqnarray}
S_{ab}(t)&=&s(\frac{M_B}{\sqrt{2}}x_1,b_1)+s(\frac{M_B}{\sqrt{2}}x_3(1-r^2),b_3)+s(\frac{M_B}{\sqrt{2}}(1-x_3)(1-r^2),b_3)\nonumber\\&&
+\frac{5}{3}\int_{1/b_1}^t\frac{d\mu}{\mu}\gamma_q(\mu)+2\int_{1/b_3}^t\frac{d\mu}{\mu}\gamma_q(\mu),\nonumber\\
S_{cd}(t)&=&s(\frac{M_B}{\sqrt{2}}x_1,b_1)+s(\frac{M_B}{\sqrt{2}}x_2,b_2)+s(\frac{M_B}{\sqrt{2}}(1-x_2),b_2)\nonumber\\&&
+s(\frac{M_B}{\sqrt{2}}x_3(1-r^2),b_1)+s(\frac{M_B}{\sqrt{2}}(1-x_3)(1-r^2),b_1)
\nonumber\\&&+\frac{11}{3}\int_{1/b_1}^t\frac{d\mu}{\mu}\gamma_q(\mu)+2\int_{1/b_2}^t\frac{d\mu}{\mu}\gamma_q(\mu),
\end{eqnarray}
where the function $s(Q,b)$ is given in \cite{epjc11695}. $\gamma_q=-\alpha_s/\pi$ is the anomalous dimension of the quark.
The threshold resummation factor $S_t(x)$ is adopted
from \cite{prd65014007},
 \begin{eqnarray}
S_t(x)=\frac{2^{1+2c}\Gamma (3/2+c)}{\sqrt{\pi}\Gamma(1+c)}[x(1-x)]^c,
\end{eqnarray}
with a running  parameter $c(Q^2)=0.04Q^2-0.51Q+1.87$ \cite{prd80074024}
and $Q^2=M^2(1-r^2)$ \cite{prd91094024}.

\section{THE WAVE FUNCTIONS}\label{sec:wave}

In the PQCD approach, the necessary inputs contain
the light-cone distribution amplitudes (LCDAs) which are constructed by  the nonlocal matrix elements.
The $B_{u,d,s}$ meson light-cone matrix element are decomposed into the following two Lorentz structures \cite{bbns}:
\begin{eqnarray}
\int d^4 z e^{ik_1 \cdot z}\langle 0|q_{\alpha}(z)\bar{b}(0)_{\beta}|B_q(P_1)\rangle=\frac{i}{\sqrt{2N_c}}
\{(\rlap{/}{P_1}+M)\gamma_5[\Phi_{B_q}(k_1)-\frac{\rlap{/}{n}-\rlap{/}{v}}{\sqrt{2}}
\bar{\Phi}_{B_q}(k_1)]\}_{\alpha\beta},
\end{eqnarray}
with  the color factor $N_c$.
As usual  the former Lorentz structure in  above equation is the dominant contribution
in the numerical calculations, while the latter Lorentz structure is negligible \cite{epjc28515}.
In impact coordinate space the $B$ meson wave function can be expressed by \cite{prd65014007,prd63054008}
\begin{eqnarray}
\Phi_{B}(x,b)=\frac{i}{\sqrt{2N_c}}(\rlap{/}{P_1}+M)\gamma_5\phi_{B}(x,b),
\end{eqnarray}
where $b$ is the conjugate variable of the transverse
momentum of the valence quark of the meson. The distribution amplitude $\phi_B(x,b)$ as being used
in Refs.~\cite{prd65014007,prd76074018} are adopted here
\begin{eqnarray}
\phi_{B}(x,b)=N x^2(1-x)^2\exp[-\frac{x^2M^2}{2\omega^2_b}-\frac{\omega^2_bb^2}{2}],
\end{eqnarray}
with the shape parameter $\omega_b$ and the normalization constant
$N$ being related to the decay constant $f_B$ by normalization:
\begin{eqnarray}
\int_0^1\phi_{B}(x,b=0)d x=\frac{f_{B}}{2\sqrt{2N_c}}.
\end{eqnarray}
The shape parameter $\omega_b=0.40\pm 0.04$ GeV for  the $B_{u,d}$ mesons and
$\omega_b=0.50\pm0.05$ GeV for the $B_s$ meson.

 For the $\psi(2S)$ meson,
 the longitudinally and transversely polarized  LCDAs up to twist-3   are defined by \cite{zhourui1,zhourui2}
\begin{eqnarray}
\langle \psi(2S) (P_2, \epsilon^L_2)|\bar{c}(z)_{\alpha}c(0)_{\beta}|0\rangle &=& \frac{1}{\sqrt{2N_c}}\int_0^1 dxe^{ixP_2\cdot z}
[m_{\psi(2S)}\rlap{/}{\epsilon^L_2}_{\alpha\beta}\psi^L(x,b)+(\rlap{/}{\epsilon^L_2}\rlap{/}{P_2})_{\alpha\beta}\psi^t(x,b)], \nonumber\\
\langle \psi(2S) (P_2, \epsilon^T_2)|\bar{c}(z)_{\alpha}c(0)_{\beta}|0\rangle &=& \frac{1}{\sqrt{2N_c}}\int_0^1 dxe^{ixP_2\cdot z}
[m_{\psi(2S)}\rlap{/}{\epsilon^T_2}_{\alpha\beta}\psi^V(x,b)+(\rlap{/}{\epsilon^T_2}\rlap{/}{P_2})_{\alpha\beta}\psi^T(x,b)].
\end{eqnarray}
 The asymptotic models  for the twist-2 distribution amplitudes
$\psi^{L,T}$ and the twist-3 distribution amplitudes $\psi^{V,t}$ are  extracted from the correspond Schr$\ddot{o}$dinger states
for the harmonic-oscillator potential. Their expressions  have been derived as \cite{zhourui1}.
\begin{eqnarray}\label{eq:wave}
\psi^{L,T}(x,b)&=&\frac{f_{\psi}}{2\sqrt{2N_c}}N^{L,T} x\bar{x}\mathcal {T}(x)
e^{-x\bar{x}\frac{m_c}{\omega_c}[\omega^2_cb^2+(\frac{x-\bar{x}}{2x\bar{x}})^2]},\nonumber\\
\psi^t(x,b)&=&\frac{f_{\psi}}{2\sqrt{2N_c}}N^t (x-\bar{x})^2\mathcal {T}(x)
e^{-x\bar{x}\frac{m_c}{\omega_c}[\omega^2_cb^2+(\frac{x-\bar{x}}{2x\bar{x}})^2]},\nonumber\\
\psi^V(x,b)&=&\frac{f_{\psi}}{2\sqrt{2N_c}}N^V [1+(x-\bar{x})^2]\mathcal {T}(x)
e^{-x\bar{x}\frac{m_c}{\omega_c}[\omega_c^2b^2+(\frac{x-\bar{x}}{2x\bar{x}})^2]},\nonumber\\
\end{eqnarray}
with
\begin{eqnarray}\label{eq:wc}
\mathcal {T}(x)=1-4b^2m_c\omega_c x\bar{x}+\frac{m_c(x-\bar{x})^2}{\omega_c x\bar{x}},
\end{eqnarray}
where the parameter $\omega_c=0.20\pm 0.02$ GeV.
$N^i(i=L,T,t,V)$  are the normalization constants and the normalization conditions:
\begin{eqnarray}
\int_0^1\psi^{i}(x,0)d x&=&\frac{f_{\psi}}{2\sqrt{2N_c}}.
\end{eqnarray}
For a light vector meson, the light-cone wave function for longitudinal (L) and
transverse (T) polarization are written as \cite{LCDAs}
\begin{eqnarray}
\Phi_V^L(x_3)&=&\frac{1}{\sqrt{2N_c}}[m_V\rlap{/}{\epsilon^L_3}\phi_V(x_3)
+\rlap{/}{\epsilon^L_3}\rlap{/}{P_3}\phi_V^t(x_3)+m_V\phi_V^s(x_3)],\nonumber\\
\Phi_V^T(x_3)&=&\frac{1}{\sqrt{2N_c}}[m_V\rlap{/}{\epsilon^T_3}\phi_V^v(x_3)
+\rlap{/}{\epsilon^T_3}\rlap{/}{P_3}\phi_V^T(x_3)+im_V\epsilon_{\mu\nu\rho\sigma}
\gamma_5\gamma^{\mu}\epsilon^{T\nu}_3v^{\rho}n^{\sigma}\phi_V^a(x_3)],
\end{eqnarray}
respectively, where $\epsilon_{0123}=1$ in our convention. Note that $v$ is the moving direction of vector particle.
The twist-2 distribution amplitudes are given by
\begin{eqnarray}\label{eq:twist2}
\phi_V(x)&=&\frac{f_V}{\sqrt{2N_c}}3x(1-x)[1+a_{1V}^{\parallel}3t+a_{2V}^{\parallel}3(5t^2-1)/2],\nonumber\\
\phi_V^T(x)&=&\frac{f_V^T}{\sqrt{2N_c}}3x(1-x)[1+a_{1V}^{\perp}3t+a_{2V}^{\perp}3(5t^2-1)/2],
\end{eqnarray}
 and those of twist-3 ones are
\begin{eqnarray}
\phi_V^t(x)&=&\frac{3f_V^T}{2\sqrt{2N_c}}t^2,\quad \phi_V^s(x)=-\frac{3f_V^T}{2\sqrt{2N_c}}t,\nonumber\\
\phi_V^v(x)&=&\frac{3f_V}{8\sqrt{2N_c}}(1+t^2),\quad \phi_V^a(x)=-\frac{3f_V}{4\sqrt{2N_c}}t,
\end{eqnarray}
with $t=2x-1$. The  vector (tensor) decay constants $f_V(f_V^T)$ together with the Gegenbauer moments \cite{jhep03069} are shown
numerically in Table \ref{tab:dc}. Note that positive $a_{1}^{\parallel,\perp}$ refer to a $\bar{K}^{*0}$ containing an $s$ quark, while for a
$K^{*+}(K^{*0})$ with an $\bar{s}$ quark, $a_{1}^{\parallel,\perp}$ changes sign \cite{prd710140029}.
\begin{table}
\caption{Input values of the decay constants  and the Gegenbauer moments \cite{jhep03069} of the light vector
mesons.}
\label{tab:dc}
\begin{tabular}[t]{lcccccc}
\hline\hline
Vector & $f_V$ (MeV)& $f_V^T (MeV)$
 & $a_{1V}^{\parallel}$ & $a_{2V}^{\parallel}$  & $a_{1V}^{\perp} $&$ a_{2V}^{\perp}$\\
\hline
$\rho$ &$216\pm 3$ &$165\pm9$ &-- & $0.15\pm 0.07$ &--&$0.14\pm 0.06$\\
$\omega$ &$187\pm5$ &$151\pm9$ &-- &$0.15\pm 0.07$ &--&$0.14\pm 0.06$\\
$K^*$ &$220\pm5$ &$185\pm10$ &$0.03\pm0.02$ & $0.11\pm0.09$ &$0.04\pm0.03$&$0.10\pm0.08$\\
$\phi$ &$215\pm5$ &$186\pm9$ &-- & $0.18\pm0.08$ &--&$0.14\pm0.07$\\
\hline\hline
\end{tabular}
\end{table}
\end{appendix}

\end{document}